\documentclass[3p]{elsarticle}

\usepackage{amsmath}

\usepackage{afterpage}

\usepackage{color, soul} 

\usepackage{adjustbox}
\usepackage{hhline}
\usepackage{caption}

\usepackage{lineno,hyperref}
\modulolinenumbers[5]

\newif\ifplotbitmap
\plotbitmaptrue

\begin{document}

\begin{frontmatter}

\title{A reflexion on the potential evolution of the CLS machine complex : A "Gedanken" experiment }


\author{F. Le Pimpec\corref{mycorrespondingauthor}}
\cortext[mycorrespondingauthor]{Corresponding author}
\ead{frederic.le.pimpec@lightsource.ca}
\author{C. Baribeau, T. Batten, G.  Bilbrough, B. Fogal, M. Ratzlaff, H. Shaker, M. Sigrist, X.~Shen, X. Stragier, W. Wurtz}
\address{Canadian Light Source Inc., University of Saskatchewan, Saskatoon, Saskatchewan, S7N 2V3, Canada}

\begin{abstract}
In the next 5 to 10 years, new 4$^{th}$ generation light sources will be coming online either from upgrades or as brand new facilities. The question regarding the competitiveness and the usability of a 3$^{rd}$ generation light source is certainly in the mind of the scientific personnel, which includes both the beamline and the accelerator staff of any 3$^{rd}$ Gen Research Infrastructures. At the Canadian Light Source (CLS), Accelerator Operation and Development (AOD) staff are responsible for the daily operation of the machines, their upgrades which extend to long term development and shall also raise awareness on their possibilities.
The boundaries of this mind experiment are 1) we must use CLS infrastructure, 2) identify the existence of the knowledge to realize an idea, 3) the realization, with whom, how, with which funds and under which administrative supervision is out of bounds.
\end{abstract}

\begin{keyword}
Storage Ring \sep CLS \sep Infrastructure 
\end{keyword}

\end{frontmatter}


\section{\label{sec:intro}Introduction}

In order to meet Canada’s post-pandemic economic and environmental goals, it is imperative that Canadian academic and industrial researchers have access to the next generation of analytical tools. Science and knowledge advancement are also dependent of those advanced tools to validate theories, discover, or understand, through sometimes serendipity, how nature works. The Canadian Light Source is a 3$^{rd}$ generation light source dating since 2004, \cite{CLSReport:2001-2004}. It is the most advanced synchrotron facility that Canada has and serves many communities, from Physics, to chemistry, material, biology, agricultural science etc.
However, the facility is aging and independently of being a formidable ally that enables discovery, needs its upgrade. This can be through a massive overhaul of its current Storage Ring (SR), transforming it from a 3$^{rd}$ generation light source to a 4$^{th}$ generation synchrotron light source, also called diffraction limited storage rings \cite{Eriksson2014a}. It can be through an enhancement of the capacity of the existing beamlines, eventually transforming them to respond to the user community demand, Canadian and international. It can also be through adding new smaller ancillary system, like Secondary Electron Microscope (SEM) (environmental, cryo etc.), more advanced biology or physics/chemistry lab, or by proposing new offers to the scientific community based on what the Accelerator Operation and Development (AOD) department could imagine using its existing infrastructure, and equipment.

In what follows we will not address science, technology and innovation that can be linked to beamline upgrades nor what a 4$^{th}$ generation light source, either a green field Free electron lasers (FEL) or a diffraction limited storage rings could provide in terms of opportunities. Looking ahead on how to support the future of CLS and using the CLS existing infrastructures and boundaries, the accelerator physicists, prompted by questions from user scientists have looked at possibilities offered by the existing accelerators within their enclosures. Possibilities do not imply feasibility, but they shall serve as a stepping stone for broader discussion among the CLS user community and beyond extending to the Canadian scientific community. \newline
We will, only, look at what the CLS facility, on the accelerator side, with its current capacity in terms of human knowledge could provide independently of the cost or resources needed to realize an idea. This exercise also provides a status quo of the actual AOD team in terms of a forward looking development.


\section{CLS a Facility Centered around its Third Generation Light Source and evolving beyond it}

As already mentioned, in the next 5 to 10 years, new 4$^{th}$ generation light sources, either linac based like Free electron lasers (FEL) or diffraction limited storage rings, will be coming online. Those diffraction limited storage ring facilities are the results of planned upgrades of existing synchrotron light sources and in some cases because an upgrade to the existing ring is not possible usually due to the existing space in the existing tunnel enclosure, or because one may want to repurpose an older machine, nor is cost efficient, are hence new light sources. The question regarding the competitiveness and the usability of a third generation light source is certainly in the mind of the scientific personnel, which includes both the beamline and the accelerator staff, of any 3$^{rd}$ generation light source. At CLS, the staff from the AOD department are responsible for the daily operation of the machines, their technical upgrades, which extend to long term development of subcomponents. AOD mandate is also to raise awareness on the possibilities of further advance and implementation of new beam techniques for the current storage rings. When looking at the evolution of equivalent research infrastructure that are centered around their storage rings, one has seen the development of advanced laboratory to prepare samples, eventually being requested by the user community (academia or industry) to prepare and qualify samples without the request of using the light source to conduct an experiment. The light sources are also moving forth with new techniques that are not requesting the use of the light but of electrons for example by purchasing Cryo-Electron Microscopes \cite{adrian1984},\cite{ESRF-CM01}. This diversification becomes necessary as the overall technologies in many domains are progressing rendering some equipment or instrument obsolete. The Canadian Light Source in order to retain its user community has to think about the diversification of its services that goes beyond the light source itself \cite{CLS-SEM}.

\section{The CLS Machine: Today}
The CLS machine provides a 2.9~GeV electron beam circulating in a 171~m circumference ring. The light produced in magnetic bending dipoles and in insertion devices, undulators or wigglers, cover the light spectrum from Far Infrared ($300~GHz <\nu<20~THz$) to hard X-Rays ($\sim 0.01~nm < \lambda <10~nm $). By lowering the electron beam energy, one can shift the wavelength spectrum to lower wavelength with more power for the same current. The electron beam loses energy in its bend magnets proportionally to its energy by the power of fourth $\sim$ E$^4$. This lost energy is rejuvenated by the RF power system that compensate the loss per turn of the beam. Currently with all photon radiators in use (Magnetic bends, and insertion devices), CLS must compensate for 248~kW with 220~mA of circulating current.

\section{The "Gedanken" experiments}
\subsection{Lowering the Energy of the Storage Ring}
\begin{figure}
\centering
\resizebox{0.75\textwidth}{!}{%
  \includegraphics{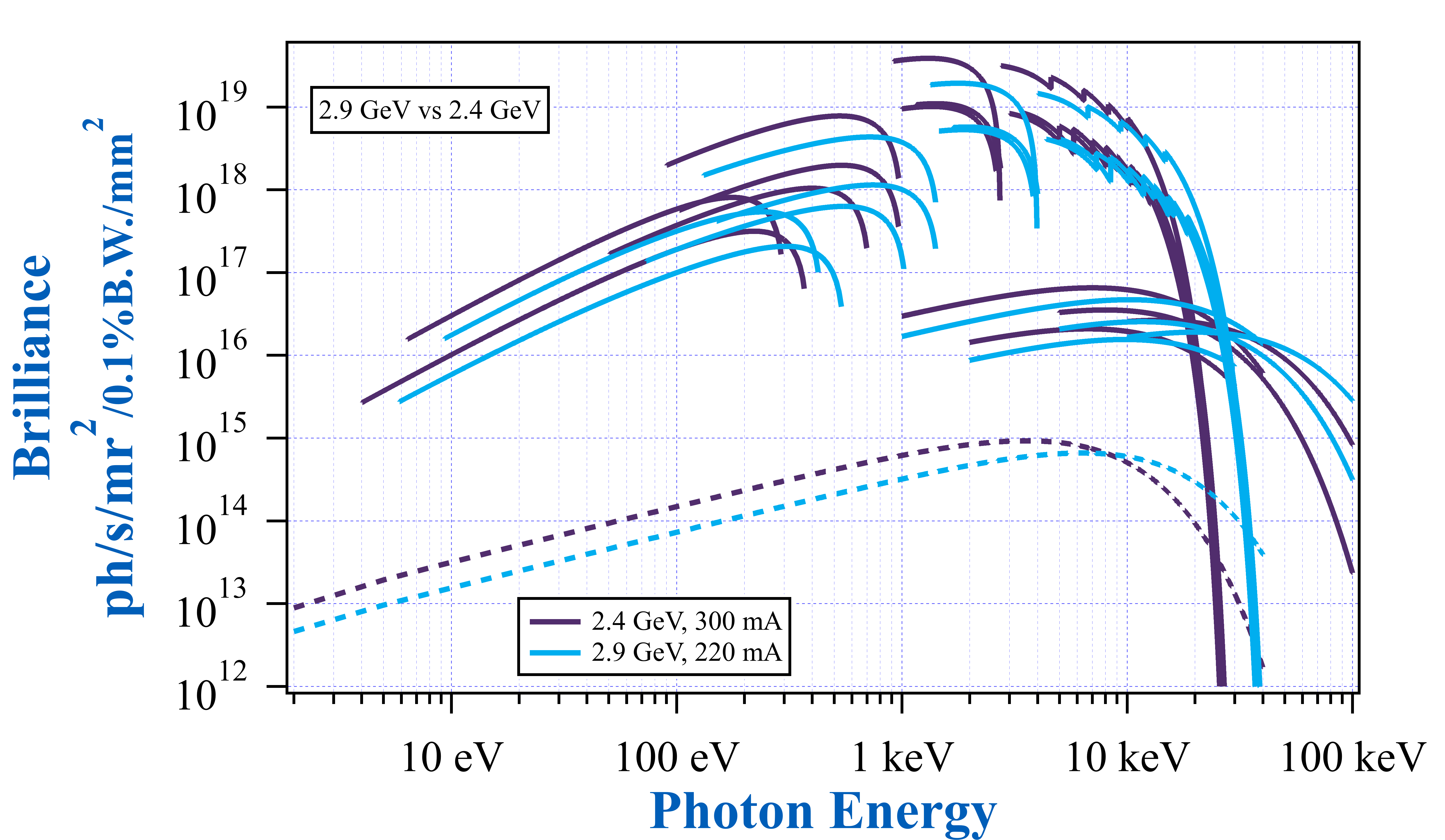}
}
\caption{Photon beam Brilliance comparison of CLS operating at two different setpoints, energy and circulating current}
\label{flux-Magnets}       
\end{figure}

By lowering the beam energy one has readily available an excess of RF power. This excess can be used to increase the beam current circulating in the ring; for 300~mA of circulating electron beam at 2.4~GeV, the power load is 232~kW. The number of photon emitted increases linearly with the current in the ring. This trick can be played until the beam energy becomes too low and the electron beam becomes unstable due to intra-beam collision (Touschek effect). A comparison for CLS of two different storage ring (SR) setpoints and their impact on the photon brilliance spectrum from bend magnets and insertion devices (undulators and wigglers) is shown in Figure.\ref{flux-Magnets}.

\subsection{Emittance swap-out}
Creating a round beam rather than using the flat beam by emittance exchange (EEX), in our case transverse emittance swap-out (vertical-Horizontal) can also be interesting for the scientists \cite{Aiba:IPAC2015-TUPJE045, Kuske:IPAC2016-WEOAA01, Aiba-PRAB-23-020701}. This requires that the ring vertical dynamic aperture is sufficiently big to conserve the circulating beam. It has also an advantage as it can increase the beam lifetime by diminishing the Touschek effect. The downside is that the brilliance of a round beam is less than the one of a flat beam \cite{LEE20233866}.

\subsection{Linac usage}
\begin{figure}[ht]
\centering
\resizebox{0.75\textwidth}{!}{%
  \includegraphics{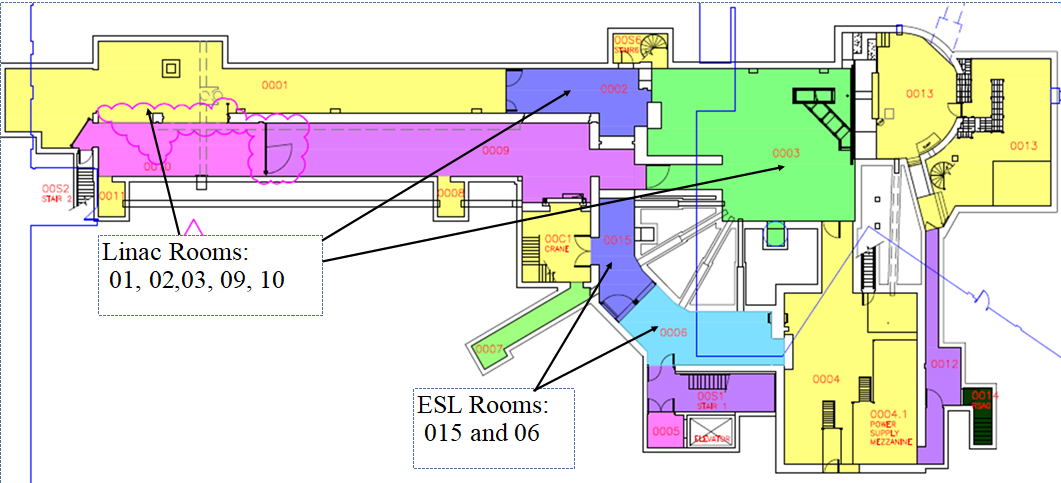}
}
\caption{CLS linac tunnel with linac installed in room 01 and 02 a transfer line in room 3 and 4 and a possibility to branch out in room 3 to use room 09 for a short FEL or THz source}
\label{linacTunnel}   
\end{figure}

For some applications and using the magnetic chicane in a linac (FEL for examples) transverse to longitudinal emittance exchange can be of interest to produce very short pulses \cite{Ha:2019qap}. CLS could explore this avenue for an electron source producing sub-picosecond long bunches, that could be used to produce intense Infra-Red (IR) or THz radiation. One would take advantage of the two tunnel sections to install the needed radiator (insertion device or a bend magnet), Figure.\ref{linacTunnel}.

The new CLS linac built and installed by Research Instrument GmbH is compact and shall, due to license condition, operate at 1~Hz but can technically operate at 10~Hz. Its main goal is to be an injector for the Storage Ring for top-up operation \cite{Boland:2023yli,Lepimpec:2025Linac}. One shot every 3 to 5 minutes is necessary to ensure smooth and continuous operation of the SR with a circulating beam around 220~mA. In between those shots the linac idles and could be used for other purposes. As previously mentioned, it could produce, with the addition of the appropriate radiator (Insertion device, bend magnet, Smith Purcell grating) intense and coherent IR or THz radiation \cite{fukanaga2010, yu2012}. If a laser was available, one could collide the photons coming from an intense laser source with the electron beam of the linac and produce an inverse Compton radiation source in up to very hard X-Ray regime \cite{DANGELO20001,PhysRevAccelBeams.28.023401}.
\newline The short bunches (~ps long), or shorter produced by a specially designed magnetic bunch compressor, could also be used to enable a new field of research at CLS to access ultra-short phenomena by using the electrons to do Ultra-fast Electron Diffraction (UED) \cite{Kim2020-gj, Centurion2022-kb}.
\newline The linac injector could also be used to explore and built or strengthen the knowledge in Canada for cancer therapy using Flash Therapy \cite{Vozenin2022-xu, Liu2023-pw}.

\subsection{Electron source and Magnet laboratories}
Adjacent to the linac, the CLS has a location reserved for an electron source laboratory (ESL), see Figure.\ref{linacTunnel}. A certain number of electron guns are at the disposal of the CLS team, including an RF photogun. Such laboratory and the readily available electron source, laser based followed by a short 50~cm long RF structure can provide a tunable electron beam of up to 10~MeV. Such setup complemented by a bend dipole magnet can be used to produce THz radiation. Taming the radiation for useful application is another subject that involves knowledge beyond the one the accelerator team possesses. The electron source installed in the ESL could be used to calibrate detectors that beamline scientists would use at their beamlines by producing various wavelength and if broadband a monochromator could be used to choose the wavelength required.

The existing accelerator team is proficient in magnet designs and characterization. A magnet lab exists that is open beyond CLS to characterize magnetic fields of various type of magnets, this certainly could include pulsed-magnets that could be used for an end-station at the SR, for example the pulsed magnet used at the High Energy Density Scientific Instrument at the European XFEL \cite{Zastrau2021-jj}.

\subsection{The Rings}
One of the critical piece of equipment at the CLS is the booster synchrotron ring (BR) that captures the 150~MeV to 250~MeV electrons coming from the linac and accelerate them to 2.9~GeV with a repetition rate of 1~Hz. The BR is now 20 years old and new upgrade are necessary, including upgrading the old power supplies to provide better stability. An upgrade of the BR with different injection scheme, on-axis versus off-axis, an improved dynamic aperture and better emittance would be the prerequisite for an improvement of the SR. However, the characteristic of such design would depend of the characteristic of the SR in which the electrons would be injected; without such understanding of the SR pre-requisite the work on a BR will stay academic. Yet, and like the linac the BR idles for long minutes. One could certainly imagine, like the Proton Synchrotron of CERN, a multi-function accelerator which primary tasks is the injection in the SR for top-up. One function could be to extract the beam at an energy below 2.9~GeV if one wanted to and send it into a suite of undulators where a laser source could modulate the electron beam to provide a FEL type beam, the wavelength range could be VUV to soft X-Ray as to produce a photon beam that is intense enough to be of interest. This would require a bit of distance to line up the insertion devices. As for the linac the BR beam after extraction (at various energy) could be collided with an intense laser to produce some inverse Compton sources. Given the configuration of the CLS Infrastructure around the BR, for both the FEL or the inverse Compton sources, it will certainly necessitate a long transfer line for the electrons to go to a dedicated experimental hall to collide with an intense laser beam or pass on dedicated undulators for FEL radiation production.

As we discussed a potential upgrade of the BR and its parameters, the necessary condition to work on the BR is to modify the lattice of the SR and transform the double bend achromat lattice into a Multi-bend achromat (MBA) lattice \cite{Dallin2016} with a set of machine and beam parameters. Modern 4$^{th}$ generation Synchrotron lattice uses 5 to 9 bend achromats. There is little room in the present space to have such an extensive lattice that is easily serviceable. One shall consider that with new technologies in permanent magnets including the possibility to build combined function magnets, nothing precludes the possibility to move beyond a double bend achromat. One may look at a triple bend Achromat, the use of super bends, and a compact lattice, using permanent magnets. One has to realize that CLS is a real synchrotron used as a SR, it could as we mentioned already operate at different electron beam energy. The use of permanent magnets freezes the lattice and the energy of the electron beam. The other challenge of a MBA lattice design is to ensure the location of the 22~beamlines. Many beam lines are fed from a bend magnet, using an MBA will for the same circumference reduce the wavelength that is reachable. The design will have to ensure compensation using superbends and antibends to ensure usability of the beamline. The other difficulty, beyond ensuring the operation of the BLs that CLS serve, is to keep as much as possible the current light source point from the machine as to not displaced the beamlines. The work is possible but requires extra knowledge that CLS does not possess anymore in lattice design.

\subsection{Other development}
One may also contemplate the novelty of Plasma wakefield acceleration, laser based (LWFA), which is also studied in Canada \cite{Kieffer2023-lv}. As knowledge advances, better control on the beam is possible that can make LWFA already today a competitor to more traditional X-Ray sources \cite{Hussein2019-rg, Svendsen-2022}. With the ESL lab and adjacent location away of the linac tunnel available, R\&D could be started to support the existing development in Canada.

\begin{figure}[!htb]
\centering
\resizebox{0.75\textwidth}{!}{%
  \includegraphics{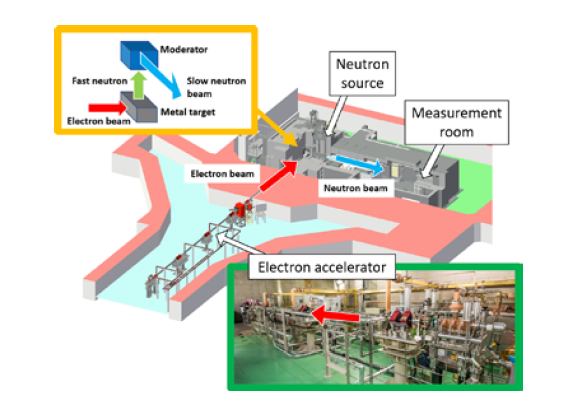}
}
\caption{Overview of the AISTANS neutron generator facility. Neutrons are generated by injection of high energy electrons, 40~MeV into a metal target. They are then moderated and extracted as a slow beam for analyzing structural materials \cite{AISTAN_NMIJ2020}}
\label{AISTANlinacTunnel}   
\end{figure}

Finally, the CLS facility could be hosting a small neutron source using its linac as a pulsed source to create neutrons. Such an analytical facility already exists in Japan \cite{KINO2019-407,AISTAN_WAO2023}, see Figure.\ref{AISTANlinacTunnel}. CLS already host the Medical Isotope Project (MIP) \cite{CLS-MIP2011, CLS-MIP2020} and knowledge could be transferred. Next to CLS, one of the most important center using neutrons exists: the Sylvia Fedoruk Canadian Centre For Nuclear Innovation \cite{Banks-2019}. 

 Synergies could be developed between the two centers to enhance the attractiveness of Saskatoon as a comprehensive science center using different probing particles for scientific and technological advancement and innovation. Such proposal although far from the core work of CLS would enhance CLS capability, expand its user community and become the sole location in Canada where a scientific proposal could request exploration using three different types of particles. Electrons with an SEM, photons with the CLS storage ring, and neutrons produced by an electron linac. A caution note is that if one imagine having a supplementary source of photons or if we were to generate neutrons, such source would probably not be very powerful, as the machine complex would be set to feed the SR. The flux of particles produced could be sufficient to enable proof of principle experiments that would open the door for the scientists to prepare stronger proposals to request beam time or instrument time at a well targeted Research Infrastructure that produces the requested particles with the required flux, brilliance, time structure and polarization.
 
 Today many light sources welcome multidisciplinary and interdisciplinary proposals to ensure their importance in the research ecosystem of their country. Scientific review panels seems now ranking highly such scientific proposals which aggregate different research groups that mix users' expertise in the use of neutron, electron or photon sources as their usual primary analytical instruments.

\section{Summary}
The 2023 AOD team, through this exercise, had started an introspection of its capacity. It is clear as years will pass that this seminal work will be outdated. Outdated by the new technologies and outdated by the mobility (gain or loss) of accelerator physicists. Nevertheless, it laid down ideas that could be technically possible at CLS to realize, some with great effort in resources and some with reasonable one. The importance of the results of this introspection is for the scientific community to realize that CLS development can be something else than the upgrade of the Storage Ring to a 4$^{th}$ generation synchrotron light source, or through the upgrades of beamlines and existing laboratories. It is to look at what an analytical research infrastructure can do beyond its traditional core business, for CLS it is the synchrotron light production. It is for the Canadian scientific community to engage with the accelerator physicists and question them on possibilities or to force them to think beyond their daily tasks and problems. A CLS expansion can certainly be done by adding on CLS premises capabilities but it is also more intangible through potential key partnership with universities or other laboratories (private or public) to develop key instrumentation. For the later and for our purpose it would be to develop or help developing electron based sources/machines to serve a requested application.

\section{Acknowledgments}
The team would like to thank T. Regier and K. Mundboth from CLS and I.~Burgess from USASK for their active questioning and participation.
This work was supported by the Canada Foundation for Innovation, the Natural Sciences and Engineering Research Council of Canada, the National Research Council Canada, the Canadian Institutes of Health Research, the Government of Saskatchewan, and the University of Saskatchewan.

\end{document}